\documentstyle[prl,aps]{revtex}

\def\be{\begin{equation}}
\def\ee{\end{equation}}
\def\ben{\begin{eqnarray}}
\def\een{\end{eqnarray}}
\newcommand{\bei}{\begin{itemize}}
\newcommand{\eei}{\end{itemize}}
\def\ra{\rangle}
\def\la{\langle}

\def \hcal{{\cal H}}
\def \dcal{{\cal D}}
\def \ecal{{\cal E}}
\def\trace{\mbox{Tr}}
\def\n{{\otimes n}}
\def\prawo{\rightarrow}
\def\lewo{\leftarrow}
\def\pl{\leftrightarrow}
\def\nadn{{1\over n}}

\begin{document}
\draft
\twocolumn

\title{Unified approach to quantum capacities:
towards  quantum noisy coding theorem}

\author{Micha\l{} Horodecki$^{1}$,
Pawe\l{} Horodecki$^{2}$ and Ryszard Horodecki
$^{1}$}

\address{$^1$ Institute of Theoretical Physics and Astrophysics,
University of Gda\'nsk, 80--952 Gda\'nsk, Poland,\\
$^2$Faculty of Applied Physics and Mathematics,
Technical University of Gda\'nsk, 80--952 Gda\'nsk, Poland
}

\maketitle

\begin{abstract}
Basing on unified approach to {\it all} kinds of quantum capacities we show that the
rate of quantum information transmission is bounded by the maximal attainable
rate  of coherent information. Moreover, we show that, if for any bipartite
state the one-way distillable entanglement is no less than coherent
information, then one obtains Shannon-like formulas for all the capacities.
The inequality also implies  that the decrease of distillable entanglement
due to mixing process does not exceed of corresponding loss information
about a system.
\end{abstract}


The challenge  for the present quantum information theory is to determine
the quantum  capacity of noisy channels
\cite{huge,Nielsen,Lloyd,BNS,Bennett_cap,BST,BKN}.  The problem is
difficult mainly for two reasons. First, according to the present knowledge,
unlike in the classical case, there are at least five
different types of quantum capacities \cite{huge}. This is because quantum
information channel can be supplemented by one- or two-way classical
channel \cite{clas}. Moreover, there are teleportation
channels \cite{Bennett_tel,Popescu},
for which a bipartite state is a resource.
The second reason is that quantum capacity exhibits a kind of nonadditivity
\cite{Shor-Smolin} that makes them extremely hard to deal with.

As one knows, the key success of classical information theory is the famous
Shannon noisy coding theorem, giving the formula
for capacity of noisy channel \cite{Cover}
\be
C=\sup_XI(X;Y)
\ee
where the supremum is taken over all  sources $X$; $I(X;Y)=H(X)+H(Y)-H(X;Y)$
is the Shannon mutual information (with $H$ being the  Shannon entropy);
$Y$ is the random variable resulting from action of the noise  to $X$. In
quantum information theory the candidate  for
the counterpart of mutual information has been found \cite{Nielsen,Lloyd}.
It is the so-called {\it coherent information} (CI). To define it, consider
two coherent informations of the bipartite state~$\varrho$
with reductions $\varrho_A$ and $\varrho_B$:
\be
I^X(\varrho)=S(\varrho_X) -S(\varrho),\quad X=A,B,
\ee
for $S(\varrho^X)-S(\varrho)\geq0$ and $I^X=0$ otherwise.
Here $S(\varrho)=-\trace \varrho\log_2 \varrho$ is the von Neumann entropy.
Then one defines the coherent information for a channel $\Lambda$ and a source
state $\sigma$ as
\be
I(\sigma,\Lambda)=I^B ((I\otimes \Lambda)(|\psi\ra\la\psi|)),
\ee
where $\psi$ is a pure state with reduction $\sigma$ (the quantity does not
depend on the choice of $\psi$).
Now, the following connection between coherent information and
a quantum capacity is known \cite{BKN}
\be
Q_\o\leq \lim_{n\rightarrow\infty}{1\over n} \sup_{\varrho_{n}}
I(\varrho_{n},\Lambda^{\otimes
n}) \equiv I^B_\o(\Lambda)
\label{ineq-Barnum}
\ee
where $Q_\o$ is the maximal number of qubits that can be reliably sent down the
channel without any supplementary classical channel.
Note that if, instead of inequality, there were {\it equality},
then we would have analogue of Shannon formula.

Unfortunately, despite a huge effort devoted to the problem
\cite{BNS,BST,BKN}, the equality
 has not been proven so far. Moreover, remarkably, the similar inequality
is not known for other capacities
than $Q_\o$, i.e. the ones attainable at the support of
backword ($Q_\lewo$) or two-way ($Q_\pl$) classical communication
\cite{forward}. There are also capacities of quantum
teleportation channels, where the resource is bipartite state
rather than  channel.
In the latter case, transmission  requires prior manipulations
(called distillation \cite{Bennett_pur,huge}) over the shared
pairs, transforming them into  pairs in pure maximally entangled
states.
Then the quantum information can be transmitted by using
teleportation \cite{Bennett_tel}.
The manipulations include any local actions, and one- or two-way classical
communication. Correspondingly, we have  two kinds of
one-way distillable entanglement of a state $\varrho$, $D_{\rightarrow}$
or $D_{\leftarrow}$ (since $\varrho$
need not be symmetric, one distinguishes directions of classical
communication) and two-way distillable entanglement $D_\pl$.
In fact, $Q_\pl$ also necessarily involves the distillation and teleportation
process.

The latter processes are so exotic from the point of view of
classical information theory, that no analogue of theory of error correcting
codes has been worked out for them so far. In contrast, in the case
of the  capacity $Q_\o$, there exists a huge theory of quantum
codes, being a generalization of classical error correcting codes theory
\cite{codes}.

In the above context the basic question arise: {\it Is there
possible  a consistent approach  for all of the capacities}? In particular:
{\it Is there a single counterpart of  the Shannon mutual information}?
In this paper, we provide a {\it unified} framework  for {\it all}
capacities. We show
that one and the same coherent information,  although in different contexts,
is a basic quantity in each case. More specifically, we show that the
inequality (\ref{ineq-Barnum}) is, in a sense, {\it universal}.
Given any type of supplementary resources, the
maximal rate of quantum communication (quantum capacity) is bounded by
the maximal rate of coherent information attainable via
these resources (CI capacity).

Now, there remains
the fundamental question: Are the quantum capacities equal to corresponding
CI capacities? We will show
that the following hypothetical inequality (call it {\it hashing
inequality} \cite{hashing})
\be
D_{\rightarrow}(\varrho) \geq I^B(\varrho),
\label{ineq-hashing}
\ee
if satisfied for all bipartite states $\varrho$,
implies that the capacities are equal to one another in all cases
(see Theorem~2).
In other words, the hashing inequality implies the
Shannon-like formulas for quantum capacities, providing the quantum
noisy coding theorem. Consequently, we argue that
to prove (or disprove) this inequality is one of
fundamental tasks of the present information theory.
In particular, if the inequality holds, then to evaluate $Q_\o$ one would need
to consider only the maximization problem
of the right-hand-side of the inequality (\ref{ineq-Barnum}).
Finally, we show that the power of the above inequality is even more
surprising. Namely, it also implies the relation between loss of classical
information, and loss of distillable entanglement \cite{Wilkens}.

Surprisingly, the reasoning leading to our results on capacities is
extremely simple. Namely, the capacity of a channel or bipartite state
at given supplementary resource is the optimal rate
of reliable transmission of qubits.
However, this is equivalent to optimal rate of reliable sharing
two-qubit pairs in maximally entangled (in short, singlet) states \cite{huge}.
Thus, we
can imagine that Alice and Bob, started with large number $n$ of pairs
in initial
state $\varrho^\n$ (or disposed $n$ uses of quantum channel $\Lambda$),
aim to share the maximal attainable number $k$ of singlet pairs.
The capacity is just the optimal rate $k/n$.
Then, what is the coherent information $I^X_{out}$ of the output
of such protocol attaining capacity? Since the coherent information is additive,
and for singlet state $I^X=1$, then $I^X_{out}$ equals to the number $k$
of final singlet pairs. Thus the obtained {\it rate} of the
coherent information $I_{out}^X/n$
is equal to capacity. But the {\it maximal} attainable rate of coherent
information (i.e. CI capacity) is no less than the one acheievable in some
particular protocol, so that it is no less than the capacity.

Assume  now, that for any bipartite state, its capacity (i.e. distillation rate)
is no less than its
coherent information. Then we consider the following protocol. Alice and Bob
start from $k$ groups of $n$ pairs (or divide $kn$ uses
of channel into $k$ groups), with $k,n$ being large, and for any group obtain
the final state $\varrho$ of maximal attainable coherent information. Then they
distill the latter state, and, by assumption, obtain the final number
of singlet pair no less than the coherent information of $\varrho$. Since
the latter was maximal attainable one, we conclude that
the capacity of the input state $\varrho$ or channel $\Lambda$ is no
less than CI capacity. Therefore, due to previous  paragraph, the quantities
must be equal.

The presented argumentation is very intuitive. It is similar to the approach
of Ref. \cite{miary}, that has already appeared to be fruitful in a different
context \cite{irrev}. The  rigourous version of the above heuristic approach
is more or less immediate. Indeed, the main
simplification we made was the assumption that {\it exact} singlets are produced.
In fact, they are always impure (the impurity vanishes in the asymptotic
or ``thermodynamic'' limit of infinite number of input pairs). However such
simplification does not lead to wrong conclusions, if only the involved
functions exhibit suitable continuity. In the rigorous proofs below we will
use continuity  of coherent information.

Let us now pass to the rigorous part of the paper. As mentioned, we will be
concerned with four supplementary
resources $C\in\{{
\rightarrow,\leftarrow,\leftrightarrow,\o}\}$. (The last one symbolizes no
supplementary resource).
If Alice and Bob
dispose one use of a channel $\Lambda$ (directed, by convention, from Alice
to Bob) and
the supplemetary resources symbolized by $C$, then they can share
a bipartite state $\varrho$. An operation that produced in this way the state
$\varrho$ from $\Lambda$ will be denoted by $\ecal_C$ so that
\be
\varrho=\ecal_C(\Lambda).
\ee
If Alice and Bob share initially a bipartite
state $\varrho_{in}$, then we will use notation $\dcal_C$
\be
\varrho_{out}=\dcal_C(\varrho_{in})
\ee
(The letters used in our notation follows from the common associations:
usual channel capacity -- error correction, teleportation channels --
distillation).
Now, the CI capacitites are defined by
\be
I_C^X(\Lambda)=\lim_{n\rightarrow\infty}{1\over n}\sup_{\ecal_C}
I^X(\ecal_C(\Lambda^\n))
\ee
for channels, and
\be
I_C^X(\varrho)=\lim_{n\rightarrow\infty}{1\over n}\sup_{\dcal_C}
I^X(\dcal_C(\varrho^\n))
\ee
for bipartite states \cite{konsys}. Throughout the paper, the symbol $X$ stands for
$A$ or $B$.

We also define quantum capacities as follows
\cite{definitions}. Define
maximally entangled state on the space $\hcal\otimes \hcal$ by
\be
P_+(\hcal)=|\psi_+(\hcal)\ra\la\psi_+(\hcal)|,\quad
\psi_+(\hcal)={1\over \sqrt{d}}\sum_{i=1}^{d} |ii\ra,
\ee
where $|i\ra$ are basis vectors in $\hcal$, while $d=\dim\hcal$. Given a
state $\varrho$,
consider sequence of operations $\{\dcal_C^n\}$ (called protocol)
transforming the input state $\varrho^\n$ into  the state $\sigma_n$
acting on the Hilbert space $\hcal_n\otimes\hcal_n$ and
with $\dim \hcal_n=d_n$,
satisfying
\be
F_n\equiv \la\psi_+(\hcal_n)|\sigma_n| \psi_+(\hcal_n)\ra \rightarrow 1.
\ee
The asymptotic ratio attainable via given protocol is then given by
\be
D_{\{\dcal_C^n\}}(\varrho)=\lim_{n\rightarrow \infty}
 {\log_2\dim \hcal_n \over n}
\ee
Then the capacity $D_C(\varrho)$ (call it $C$-distillable entanglement)
is defined by maximum over all possible protocols
\be
D_C(\varrho)=\sup D_{\{\dcal_C^n\}}(\varrho).
\ee
The usual channel capacities can be defined in the same way.
We only need make the following substitutions: $D \prawo Q$,
$\varrho\prawo\Lambda$ and $\dcal\prawo\ecal$. The protocols $\{\ecal_C^n\}$
and $\{\dcal_C^n\}$ that achieve the considered suprema will be called optimal
error correction  and optimal distillation protocol, respectively.
The quantity $D_\o$ is a bit  pathological, but certainly
interesting quantity. We will not be concerned with it here. However, it is
likely, that $D_\o$ is the amount of pure entanglement that can be drawn from
the state reversibly.

We will need a lemma, stating that coherent information
$I^X$ is continuous on isotropic state. The latter is defined on $\hcal\otimes
\hcal$ (cf. \cite{Popescu,Werner,xor})
\be
\varrho(F,d)= p P_+(\hcal) + (1-p) {1\over d^2}I,\quad 0\leq p\leq1,
\ee
with  $\trace \big[\varrho(F,d) P_+(\hcal)\big]=F$, $d=\dim\hcal$.

{\bf Lemma.}
For a  sequence of isotropic states $\varrho(F_n,d_n)$,
such that $F_n\rightarrow 1$ and $d_n\rightarrow \infty$ we have
\be
\lim_{n\rightarrow \infty} {1\over \log_2 d_n}
I^X(\varrho(F_n,d_n))\rightarrow 1.
\ee

{\bf Proof.} This can be checked by direct calculation.

We note an important property of the isotropic
state \cite{xor}. Namely, any state $\sigma$ acting on
$\hcal\otimes \hcal$,  if
subjected to $U\otimes U^*$ twirling  (cf. \cite{Bennett_pur}), i.e. random
unitary transformations of the form $U\otimes U^*$,
becomes isotropic state $\varrho(d,F)$
with $F=\trace \big[\sigma P_+(\hcal)\big]$, $d=\dim\hcal$.

Now we can state  the theorems being the main results of
this paper.

\vskip3mm
{\bf Theorem 1.}  Quantum capacities are bounded from above by
CI capacities:
\ben
&&Q_C(\Lambda)\leq I^X_C(\Lambda),\\
&&D_C(\varrho)\leq I^X_C(\varrho)
\een
for any $\Lambda$,  $\varrho$ and
$C\in\{{
\rightarrow,\leftarrow,\leftrightarrow,\o}\}$.
\vskip3mm
{\bf Theorem 2.} If the hashing inequality
\be
D_{\prawo}(\varrho)\geq I^B(\varrho)
\ee
holds for any bipartite state $\varrho$ then the qantum capacities
are equal to corresponding CI capacities
\ben
&&Q_\pl(\Lambda)= I^X_\pl(\Lambda), \quad
Q_{\prawo\atop(\lewo)}(\Lambda)=I^{B(A)}_{\prawo\atop(\lewo)}(\Lambda),\\
&&Q_\o(\Lambda)=I_\o^B(\Lambda),\\
&&D_\pl(\varrho)= I^X_\pl(\varrho),\quad  D_{\prawo\atop(\lewo)}(\Lambda)=
I^{B(A)}_{\prawo\atop(\lewo)}(\varrho).
\een

{\bf Remarks.} (i) If we assumed ``dual'' hashing  inequality $D_\lewo\geq
I^B$, we would get the same results modulo change $A\pl B$. Our choice of
$D_\prawo\geq I^B$ is motivated by investigations of Refs.
\cite{Nielsen,Lloyd,BNS}. (ii) It follows that the hashing inequality implies
$I^A_\pl=I^B_\pl$. (iii) Our results apply to other kind of supplementary
resurces such as e.g. public bound entanglement.

{\bf Proof of Theorem 1.}
We will prove the ``Q'' part of the theorem. The proof for
``D'' part is similar.
Let $\{\ecal_C^n\}$ be the optimal error correction
protocol for $\Lambda$. Then we have the following estimates
\be
I_C^X(\Lambda)\geq {1\over n}I^X(\ecal_C^n(\Lambda^\n))\geq {1\over n}
I^X(\varrho(d_n,F_n))\prawo Q_C(\Lambda),
\ee
where $\varrho(d_n,F_n)$ is the twirled state $\sigma_n=
\ecal_C^n(\Lambda^\n)$. The first inequality comes from the very definition
of $I_C^X$. The second one follows from convexity of $I^X$ \cite{BNS},
and its invariance under product unitary tranformations (as the twirled state
is a mixture of product unitary tranformations of the initial one). Finally,
since in optimal error correction protocol we have $F_n\prawo1$, and
$\log d_n/n\prawo
Q_C$,  we obtain  the right-hand-side limit by applying continuity of $I_X$
(see lemma).

{\bf Proof of Theorem 2.}
We will also prove only the ``Q'' part. For $C\in\{ \prawo,\lewo,\pl \}$,
consider the following
particular error correcting protocol  for the channel $\Gamma=\Lambda^\n$. One
applies  to $\Gamma$ the operation $\ecal_C$ that produces the state
$\sigma=\ecal_C(\Lambda^\n)$ of maximal attainable coherent information.
Subsequently, one performs optimal distillation protocol for the state $\sigma$.
Then we find
\ben
&&Q_C(\Lambda)= \nadn Q_C(\Lambda^\n) \geq
\nadn D_C(\ecal_C(\Lambda^\n)) \geq\nonumber\\
&& \nadn I^X (\ecal_C(\Lambda^\n)) \rightarrow I_C^X(\Lambda)
\een
where $C\in\{\prawo,\lewo,\pl\}$;
$X=A,B$ for $C=\,\pl$, and $X=A\,(B)$ for $C=\,\lewo\!(\prawo)$.
The equality follows from the very definition of $Q_C$. The first inequality
comes from the fact, that $Q_C$ is supremum over {\it all} error correction
protocols, so it is no less from the rate obtained in the protocol above.
The second inequality follows from  the hashing inequality,
and from the obvious inequality $D_\pl\geq D_{\prawo\atop (\lewo)}$. Finally,
the limit is due to the definition of $I^X_C(\Lambda)$.
The above estimate together with Theorem 1 gives all the desired equalities
apart from the one involving $C=\o$. That the latter one is also implied by
the hashing inequality, it follows immediately from the facts:
(a) trivially $I^B_\o\leq I_\prawo^B$; (b) $Q_\prawo=Q_\o$ \cite{BKN}; (c)
as just proved, the hashing inequality implies $Q_\prawo=I_\prawo^B$.

Let us now prove yet another important implication of the hashing inequality.
Namely, consider the process of discarding information
\be
\{p_i,\varrho_i\}\prawo \varrho=\sum_ip_i\varrho_i.
\ee
In Ref. \cite{Wilkens} it was shown that, for a class of ensembles
$\{p_i,\varrho_i\}$, the amount  of information lost in the process is no
less than the loss of distillable entanglement $D_\pl$, and  it was conjectured
to hold in general. The loss of information is quantified by average
increase of entropy, so that the problem is whether the folowing
inequality holds
\be
\sum_ip_iD_C(\varrho_i)-D_C(\varrho)\leq S(\varrho)-\sum_ip_iS(\varrho_i).
\ee
Note, that for pure state $\psi$, $D_C(\psi)=S(\varrho^X)$, where $\varrho^X$ is
either of the reductions of $\psi$ \cite{conc}. Therefore, for pure
states $\varrho_i$, the inequality reads
\be
D_C(\varrho)\geq \sum_ip_iS(\varrho^X_i)-S(\varrho)
\ee
Applying convexity of entropy we see that the hashing inequality implies
the above one. It is interesting, that it does not seem to imply the
inequality for impure $\varrho_i$'s.

Let us list that recent results concerning entanglement distillation,
implying that it is reasonable to conjecture that the hashing inequality
(\ref{ineq-hashing}) holds. (i) In all cases where one
has sufficiently tight lower bounds for
$D_\prawo$, the inequality is known to be satisfied. For pure states, and
other ones with
entanglement of formation equal to entanglement of distillation \cite{therm}
we have $D_\prawo=I^B$. For mixtures of two-qubit Bell states we have
$D_\prawo\geq I^B$ by hashing protocol \cite{huge}. In particular, for some of
them there is equality \cite{Rains_bound}, while for other ones one has
$D_\prawo>I^B$
\cite{Shor-Smolin}. (ii) If the hashing inequality is true, then any upper bound
for $D_\pl$ should be no less than $I^X$. This was shown for entanglement
of  formation \cite{therm} and, quite recently, for relative entropy of
entanglement \cite{Plenio_coh}. We do not know yet, if the inequality holds
for the new bound for $D$ derived in \cite{irrev}. (iii) If a state is
bound entangled \cite{bound},  then  we should have $I^X=0$. It is indeed the
case. According to Ref. \cite{xor}, the  bound entangled states must
satisfy the
so called {\it reduction criterion of separability}. This implies \cite{Cerf}
that the entropic inequality $S(\varrho)\geq S(\varrho_X)$ is also satisfied,
hence $I^X=0$. Thus we see that there is a strong evidence that the
inequality is true. We believe that the present results will stimulate
to prove (or disprove) it.

The work is supported by Polish Committee for Scientific Research, contract
No. 2 P03B 103 16, and by the IST project EQUIP, contract No. IST-1999-11053.


\begin{references}
\bibitem{huge}
C. H. Bennett, D. P. Di Vincenzo, J. Smolin and
W. K. Wootters, Phys. Rev. A {\bf 54}, 3814 (1997).
\bibitem{Nielsen}
B. Schumacher, Phys. Rev. A {\bf 54}, 2614 (1996);
B. Schumacher and M. A. Nielsen, Phys. Rev. A {\bf 54}, 2629 (1996).
\bibitem{Lloyd}
S. Lloyd, Phys. Rev. A {\bf 56}, 1613 (1997).
\bibitem{BNS}
H. Barnum, M. Nielsen and
B. Schumacher Phys. Rev.  A {\bf 57}, 4153 (1998). 
\bibitem{Bennett_cap}
C. H. Bennett, D. Di Vincenzo and J. Smolin, Phys. Rev. Lett. {\bf 78}, 3217
(1997).
\bibitem{BST}
H. Barnum, J. Smolin, and B. Terhal Phys. Rev. A {\bf 58}, 3496 (1998).
\bibitem{BKN}
H. Barnum, E. Knill and M. Nielsen, quant-ph/9809010.
\bibitem{clas}
There are also {\it classical} capacities of quantum channels, see
B. Schumacher and M. Westmoreland Phys. Rev. A {\bf 56}, 131 (1997);
A. S. Holevo, IEEE Trans. IT {\bf 44}, 269 (1998);
Bennett {\it et al.} Phys. Rev. Lett. {\bf 83}, 3081 (1999);
Here we are concerned with solely  {\it quantum} capacities.
\bibitem{Bennett_tel}
C. Bennett, G. Brassard, C. Crepeau, R. Jozsa, A. Peres and W. K. Wootters,
Phys. Rev. Lett.  {\bf 70} (1993) 1895.
\bibitem{Popescu}
S. Popescu, Phys. Rev. Lett. {\bf 72}, 797 (1994).
\bibitem{Shor-Smolin}
P. Shor and J. Smolin, quant-ph/9706061; D. DiVincenzo, P. Shor and J. Smolin
Phys. Rev. A {\bf 57}, 830 (1998).
\bibitem{Cover}
T. M. Cover and J. A. Thomas, {\it Elements of Information Theory}, John
  Wiley and Sons, New York, 1991.
\bibitem{forward}
In Ref. \cite{BKN} it was shown that forward classical communication
does not increase capacity, i.e. $Q_\o=Q_{\prawo}$. The interesting question
is if $Q_{\lewo}=Q_\o$.
\bibitem{Bennett_pur}
C. H. Bennett, G. Brassard, S. Popescu, B. Schumacher, J. Smolin and
W. K. Wootters, Phys. Rev. Lett. {\bf 76}, 722 (1996).
\bibitem{codes}
P. Shor, Phys. Rev. A, {\bf 52}, 2439 (1995); A. Steane,
Phys. Rev. Lett. {\bf 77}, 793 (1996);
Th. Beth and M. Grassl, Fortschr. Phys. {\bf 46}, 459 (1998)
and references therein.
\bibitem{hashing}
For a class of states, the inequality was shown to be true  by use of
the so-called {\it hashing} distillation protocol \cite{huge}.
\bibitem{Wilkens}
J. Eisert, T. Felbinger, P. Papadopoulos, M. B. Plenio and
M. Wilkens, Phys. Rev. Lett. {\bf 84}, 1611 (2000).
\bibitem{resource}
We will assume, without explicit stating it each time, that a fixed
supplementary resource is used.
\bibitem{miary}
M. Horodecki, P. Horodecki and R. Horodecki, Phys. Rev. Lett. {\bf 84},
2014 (2000).
\bibitem{irrev}
M. Horodecki, P. Horodecki and R. Horodecki, Report No.  quant-ph/9912076.
\bibitem{konsys}
By convexity of $I^X$ \cite{BNS} this definition is consistent
with notation used in formula (\ref{ineq-Barnum}).
\bibitem{definitions}
The used definition of ``D'' capacities is due to \cite{Rains_bound} (cf.
\cite{huge}). In Ref. [E. Rains, Phys. Rev. A {\bf 60}, 173 (1999)]
the definition was shown to coincide with other used definitions.
The definition of ``Q'' capacities is taken from
\cite{Bennett_cap}. In Ref. \cite{BKN} it was shown to be equivalent
with other one introduced  in \cite{BNS}.
\bibitem{Werner}
R. F. Werner, Phys. Rev. A {\bf 40}, 4277 (1989).
\bibitem{xor}
M. Horodecki and P. Horodecki, Phys. Rev. A, {\bf 59} 4206 (1999).
\bibitem{conc}
C. H. Bennett, H. J. Bernstein, S. Popescu and B. Schumacher,
Phys. Rev. A {\bf 53}, 2046 (1996).
\bibitem{therm}
P. Horodecki, M. Horodecki and R. Horodecki, Acta Phys.
Slovaca {\bf 48}, 141 (1998).
\bibitem{Rains_bound}
E. M. Rains, Phys. Rev. A {\bf 60}, 179 (1999).
\bibitem{Plenio_coh}
M. B. Plenio, S. Virmani, P. Papadopoulos, Report No. quant-ph/0002075.
\bibitem{bound}
M. Horodecki, P. Horodecki and R. Horodecki, Phys. Rev. Lett. {\bf 80},
5239 (1998).
\bibitem{Cerf}
N. Cerf, C. Adami and R. M. Gingrich, Phys. Rev. {\bf 60}, 898 (1999).
\end{references}
\end{document}